# TRACKING ANALYSIS AND PREDICTIVE MAINTENANCE IN ORDER TO OBTAIN DYNAMICS OF MACHINE TOOL SPINDLE

## Zapciu M, K'Nevez J-Y, Gérard A.

**Abstract:** Predictive maintenance is directed towards recognizing the earliest significant changes in machinery condition. Contrasted with protective condition monitoring in which fast response is the primary requirement, predictive monitoring is not limited by time and may use a greather range of complex characteristics. Vibration analysis has long been used for the detection and identification of machine tool condition. Main focus is to identify a procedure to obtain eigenvalue frequencies for machine tool spindle using tracking analysis.

## 1. Introduction

Predictive maintenance could be accomplished on-line from installed sensors (e.g. vibration and bearing temperatures), periodically with a handled data collector or a combination of the two (fig.1).

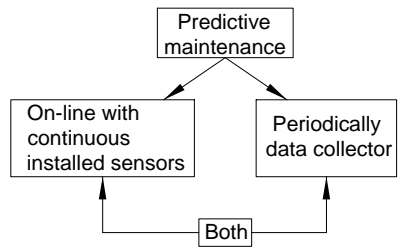

*Fig.1* Types of predictive maintenance

Comparing and trending overall vibration amplitudes, recorded at periodic intervals on selected equipment (for example data collector 2526 from

Brüel & Kjaer company) is an example of a basic predictive monitoring program.

Predictive monitoring programs have three main objectives:

1 – Earlier warning of the potential defects on equipment equipped with continuous monitoring systems;

2 – Monitoring the dynamic condition of general purpose machinery

3 – Primary warning of defects that may remain hidden within complex characteristic, for example rolling-element bearing flaws.

Most organizations evolve into a cost-effective program of predictive monitoring that includes a mixture of measurements from permanently installed transducers as well as measurements collected periodically with portable instruments.

## 2. Typical Predictive Maintenance

Most of effective predictive-monitoring programs will incorporate a wide variety of parameters to accurately characterize condition and provide earliest warning of significant changes. The process of typical predictive monitoring program, consist of four parts beginning with a detailed vibration signature on each piece of monitored equipment (Fig.2). If the component at running frequency dominates the spectrum no greater than 20 ÷ 25 % of the amplitude at running frequency, then monitoring overall amplitude is two or three broad bands is a appropriate method for early detection on rolling element bearings.

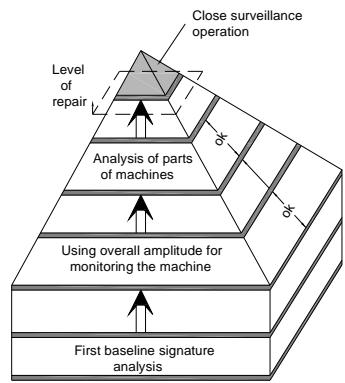

*Fig. 2* Elements of typical predictive monitoring program.

It is no coincidence that Fig.2 is a pyramid. An effective predictive monitoring program must be designed to accurately recognize the earliest significant changes and use the simplest and least costly means to separate the machine with problems from the large number of machines in good dynamic condition.

Experienced organizations report that a fully implemented of regularly predictive monitoring program will eliminate unexpected failures and reduce the number of machines in questionable marginal condition to less then 6 – 8 % of the total

### 3. Measurement of the level of spindle vibration of one high speed milling machine

In order to know the proper domain for High Speed Cutting for specific milling processes the authors propose to begin the research with the vibration of the spindle machine tool (Fig. 3) using *Tracking* analysis module of Vibroport 41 [3,4].

Eigen frequencies of the assembly spindle-bearings (no cutting process) were: 29130 rpm (first harmonic of 242.5 Hz), 14550 rpm (242.5 Hz) and 12.600 rpm (210 Hz). In the figures 4 and 5 are presented the tracking signal acquired using Vibroport 41.

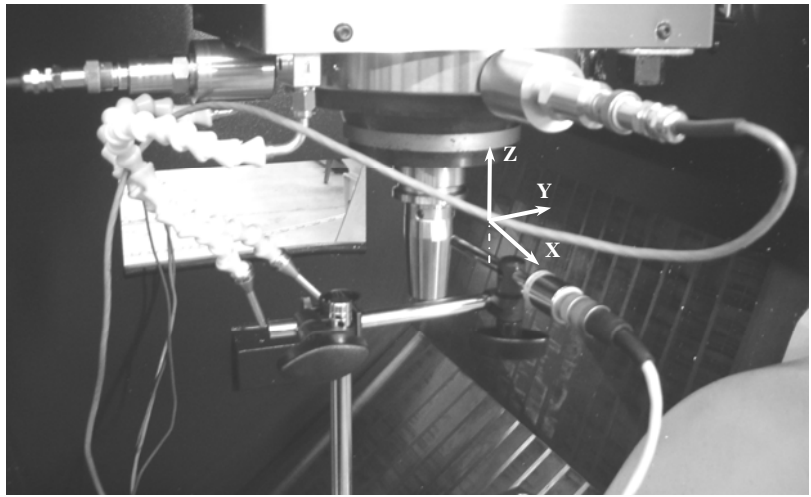

*Fig. 3* Accelerometers placement on the spindle of milling machine tool

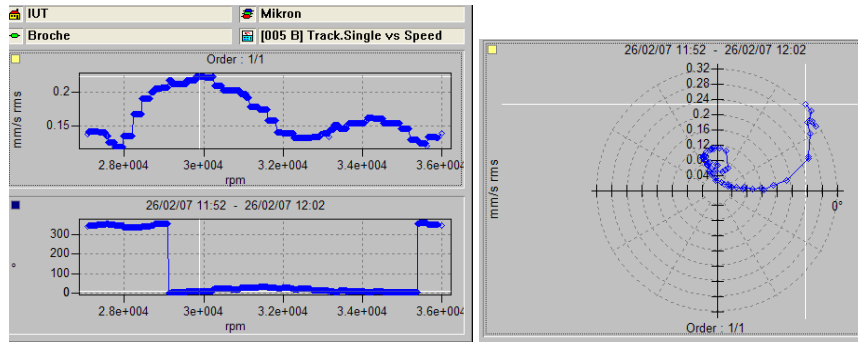

*Fig. 4* Tracking signal using speed domain 36000-28000 rpm; direction X-X

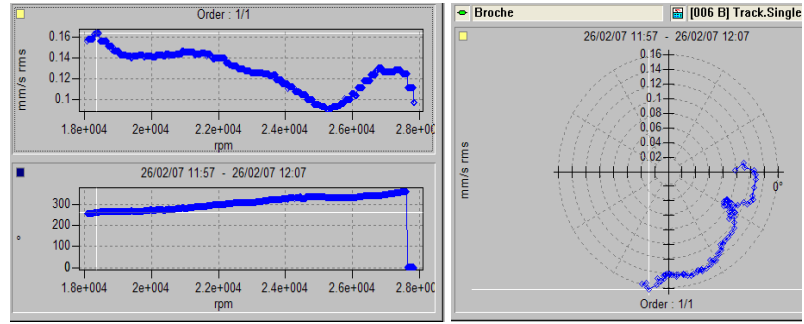

*Fig. 5* Tracking signal using speed domain 28000-18000 rpm; direction X-X

## 4. Conclusion

Main objective in this research paper was to propose a predictive maintenance for the machine tool domain. Secondly important was to separate the dynamics of the spindle of the machine tool in order to have a good control for the cutting processes. In this context, actual subject is important and it can help to elaborate a proper model for predictive maintenance. This work was validated by the experimental results based on the measuring of level of vibration of the spindle of a Milling Vertical Centre using tracking module of Vibroport 41 apparatus.

**R e f e r e n c e s**